\documentclass[footinbib,aps,pra,reprint,superscriptaddress]{revtex4-1}
\usepackage{amsmath,amssymb,braket,upgreek,bm,verbatim}
\usepackage{graphicx,tabularx} 
\usepackage{float} %
\usepackage{xcolor}

\usepackage[absolute]{textpos}
\usepackage{xcolor}
\usepackage{multirow}
\usepackage[colorlinks=true,bookmarks=false,citecolor=blue,urlcolor=blue]{hyperref}
\usepackage{lineno}
\usepackage{blindtext}
\usepackage[absolute]{textpos}
\usepackage{physics}
\usepackage{amssymb}

\begin{document}

\title{Broadband polarization-entangled source for C+L-band flex-grid quantum networks}

\author{Muneer~Alshowkan}
\email{alshowkanm@ornl.gov}
\affiliation{Quantum Information Science Section, Oak Ridge National Laboratory, Oak Ridge, Tennessee 37831, USA}
\author{Joseph~M.~Lukens}
\email{lukensjm@ornl.gov}
\affiliation{Quantum Information Science Section, Oak Ridge National Laboratory, Oak Ridge, Tennessee 37831, USA}
\author{Hsuan-Hao~Lu}
\affiliation{Quantum Information Science Section, Oak Ridge National Laboratory, Oak Ridge, Tennessee 37831, USA}
\author{Brian~T.~Kirby}
\affiliation{DEVCOM Army Research Laboratory, Adelphi, Maryland 20783, USA}
\affiliation{Tulane University, New Orleans, Louisiana 70118, USA}
\author{Brian~P.~Williams}
\affiliation{Quantum Information Science Section, Oak Ridge National Laboratory, Oak Ridge, Tennessee 37831, USA}
\author{Warren~P.~Grice}
\affiliation{Quantum Information Science Section, Oak Ridge National Laboratory, Oak Ridge, Tennessee 37831, USA}
\author{Nicholas~A.~Peters}
\affiliation{Quantum Information Science Section, Oak Ridge National Laboratory, Oak Ridge, Tennessee 37831, USA}
\date{\today}
\begin{abstract} %
The rising demand for transmission capacity in optical networks has motivated steady interest in expansion beyond the standard C-band (1530--1565~nm) into the adjacent L-band (1565--1625~nm), for an approximate doubling of capacity in a single stroke. However, in the context of quantum networking, the ability to leverage the L-band will require advanced tools for characterization and management of entanglement resources which have so far been lagging. %
In this work, we demonstrate an ultrabroadband  two-photon source integrating both C- and L-band wavelength-selective switches for complete control of spectral routing and allocation across 7.5~THz in a single setup. %
Polarization state tomography of all 150 pairs of 25~GHz-wide channels reveals an average fidelity of 0.98 and total distillable entanglement greater than 181~kebits/s. 
This source is explicitly designed for flex-grid optical networks and can facilitate optimal utilization of entanglement resources across the full C+L-band.
\end{abstract}

\maketitle

\begin{textblock}{13.3}(1.4,15)
\noindent\fontsize{7}{7}\selectfont \textcolor{black!30}{This manuscript has been co-authored by UT-Battelle, LLC, under contract DE-AC05-00OR22725 with the US Department of Energy (DOE). The US government retains and the publisher, by accepting the article for publication, acknowledges that the US government retains a nonexclusive, paid-up, irrevocable, worldwide license to publish or reproduce the published form of this manuscript, or allow others to do so, for US government purposes. DOE will provide public access to these results of federally sponsored research in accordance with the DOE Public Access Plan (http://energy.gov/downloads/doe-public-access-plan).}
\end{textblock}

Optical communications are a vital technology in the fast-growing world of data traffic, as they permit several carriers onto a single optical fiber via wavelength division multiplexing (WDM). The rapid increase in data traffic has caused a significant demand for more dense wavelength division multiplexing (DWDM) transmission capacity~\cite{Gerstel2012,Cheng2018}. However, network deployment remains an ongoing challenge due to continuous bandwidth increases. Currently, most commercially available technologies are functional in the optical C-band (1530--1565~nm). %
While reducing channel spacing is one method of increasing channel capacity, it can exacerbate nonlinear effects such as cross-phase modulation and four-wave mixing~\cite{Peters2009}. %
Utilizing the low-loss L-band  (1565--1625~nm) is an attractive alternative for increasing the number of channels within a fiber, %
which---thanks to development of L-band components such as filters, switches, pulse shapers, and wavelength-selective switches (WSSs)~\cite{Ma2020}---has become a viable choice for extending WDM capacity.

The high-fidelity distribution of entanglement between distant quantum systems is essential for realizing the potential of a future quantum internet~\cite{Wehner2018}. %
In particular, broadband polarization-entangled sources---with photons 
strongly correlated in the frequency degree of freedom (DoF) as well%
---are well positioned to utilize the full C+L-band for %
flex-grid quantum networks. The polarization DoF is a staple of quantum optics due to the ease of manipulation and measurement, while the frequency DoF can facilitate deterministic splitting of the photon pairs and various bandwidth allocation and distribution schemes. %
Such sources have been explored in various material platforms including periodically poled lithium niobate (PPLN) waveguides~\cite{Lim_2008broadband,Herbauts_2013demonstration,Vergyris2017,wengerowsky_2018entanglement,Vergyris_2019fibre,Joshi_2020trusted,Lingaraju_2021adaptive,Alshowkan2021,Alshowkan2022b,yamazaki_2022massive}, periodically poled silica fibers~\cite{Chen_2017compensation,Chen_2022telecom}, and semiconductor chips~\cite{Autebert_2016multiuser,Appas_2021flexible}. In order to characterize the quality of the entanglement, a common technique is to utilize tunable filters to select a few channel pairs across the spectrum and perform quantum state tomography (QST)~\cite{Lim_2008broadband,Chen_2017compensation,yamazaki_2022massive}. For quantum network deployment, passive optical add/drop multiplexers %
can slice the spectrum into multiple channels %
either with a single, multichannel DWDM~\cite{Herbauts_2013demonstration,Autebert_2016multiuser,Vergyris_2019fibre} or a cascade of DWDM filters~\cite{wengerowsky_2018entanglement,Joshi_2020trusted}. In addition, advanced flex-grid bandwidth allocation for broadband bandwidth have been explored based on WSSs~\cite{Lingaraju_2021adaptive, Appas_2021flexible, Chen_2022telecom,Alshowkan2021,Alshowkan2022b}. %
While most of these demonstrations have bandwidths reaching the L-band, none so far have shown the capability of managing this additional resource in a reconfigurable fashion.

In this paper, we experimentally demonstrate the management of a frequency-correlated polarization-entangled multiband photon source useful for on-demand and reconfigurable entanglement distribution. Incorporating two WSSs individually tailored to either the C- or L-band, our source contains 150~pairs of 25~GHz-wide channels spanning 7.5~THz, each aligned to the ITU grid (ITU-T Rec. G.694.1) and individually and adaptively addressable with a WSS---%
all without altering the underlying experimental setup. We perform QST of each channel, finding high fidelities (average of 0.98 over all channels) and lower-bounding the total distillable entanglement at 181~kebits/s across the full bandwidth. Our design allows entanglement management across the entire C+L-band and should prove valuable in the expansion of flex-grid quantum networks to support ever-growing numbers of users.

\begin{figure}[t!]
	\centering
    \includegraphics[width=\columnwidth]{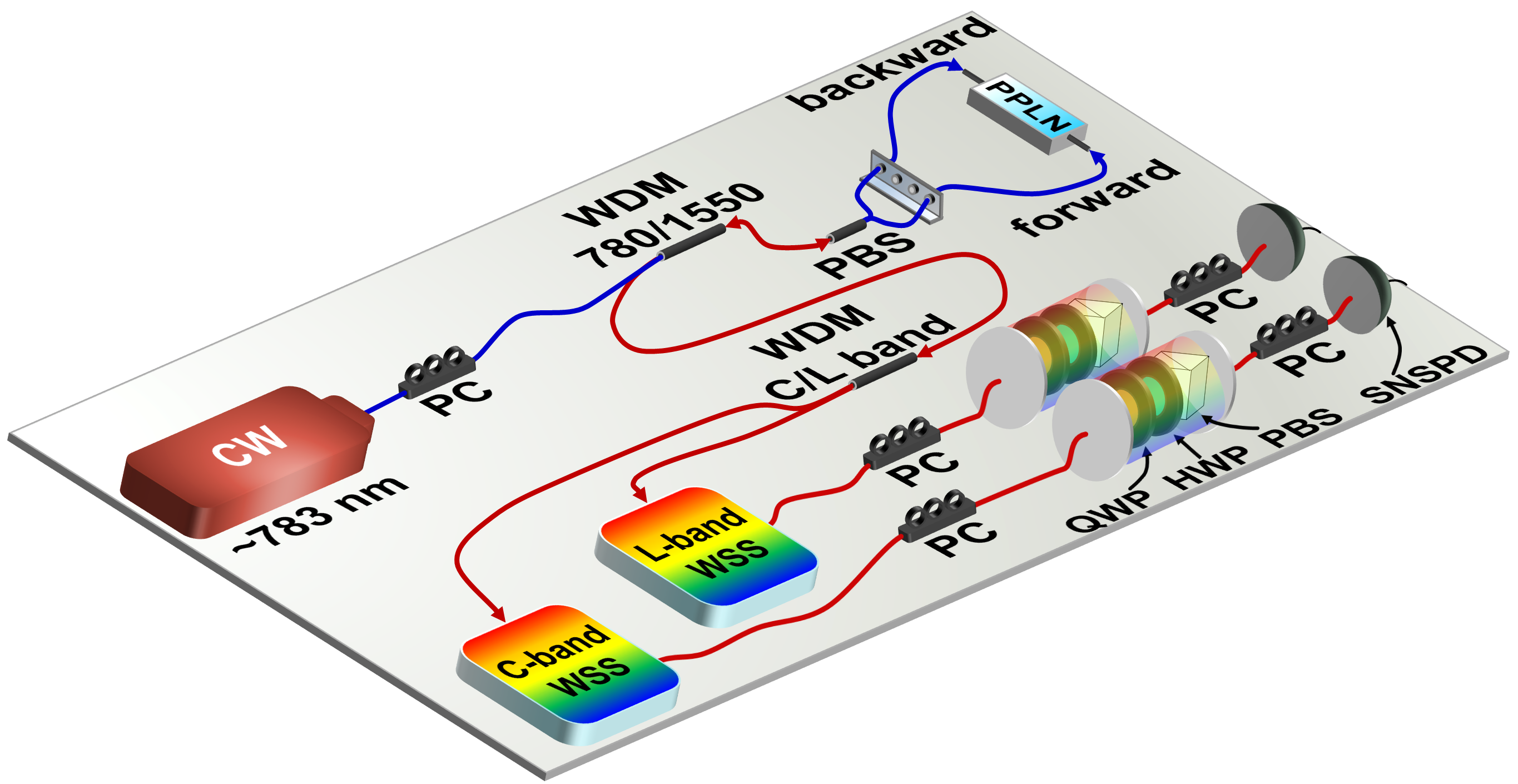}
	\caption{Experimental setup.
	Blue lines: polarization-maintaining fiber.
    Red lines: non-polarization-maintaining fiber.	CW: continuous-wave tunable laser.
	PC: fiber polarization controller.
	WDM: wavelength division multiplexer.
	PPLN: periodically poled lithium niobate waveguide.
	WSS: wavelength-selective switch.
	PBS: polarizing beamsplitter.
	HWP: half-wave plate.
	QWP: quarter-wave plate.
	SNSPD: superconducting nanowire single-photon detector.
	}
	\label{fig1png}
\end{figure}

The source consists of a fiber Sagnac loop, as shown in Fig.~\ref{fig1png}, which is based on a 12~mm-long PPLN ridge waveguide (AdvR) designed for producing---via type-0 spontaneous parametric downconversion (SPDC)---spectrally correlated, polarization-entangled photons in the ideal Bell state $\ket{ \Psi^+}\propto\ket{HH}+\ket{VV}$, following the design of~\cite{Vergyris2017}. We aim to generate biphoton bandwidth that roughly evenly fills the C- and L-band, so we tune the crystal's temperature to maximize SPDC efficiency %
from a continuous-wave pump at 383~THz (782.748~nm), producing frequency-correlated photons centered at 191.5~THz (1565.496~nm) and spanning roughly 18~THz (150~nm). The pump is connected to a  polarization controller (PC) that outputs diagonally polarized photons to a 780/1550~nm WDM. %
A fiber-based polarizing beam splitter (PBS) receives the pump and outputs the orthogonally polarized components into polarization-maintaining fibers. The fiber-pigtailed source with 980~nm polarization-maintaining fibers aligned to the slow axis %
receives and directs the pump to the waveguide in both directions (denoted ``forward'' and ``backward''). The crystal then converts the vertically polarized pump photons $\ket{V}$ to $\ket{VV}$. Although the pumps in each direction are converted in the same way, one direction is rotated again at the PBS due to the 90-degree rotated fiber. %
Consequently, polarization-entangled states produced within the waveguide exit the PBS to the WDM as $\alpha\ket{HH} + \beta\ket{VV}$, with weights determined by the incident laser polarization and any loss imbalance.

The 780/1550~nm WDM routes the biphotons to a C/L-band WDM that sends the higher frequencies to a C-band WSS (191.325--196.150~THz; Finisar) and lower frequencies to an L-band WSS (186.075--191.075~THz; Finisar). In each WSS, we section the bandwidth into 150~channels of 25~GHz bins aligned to the ITU grid; we choose 25~GHz since we found it to be the minimum channel width at which no additional peak loss was observed on the WSS filters. Because each channel is frequency-correlated with another in the complementary WSS, the slices in the C-band and the L-band span a total of 7.5~THz.

\begin{figure}[tb!]
	\centering
    \includegraphics[width=2.8in]{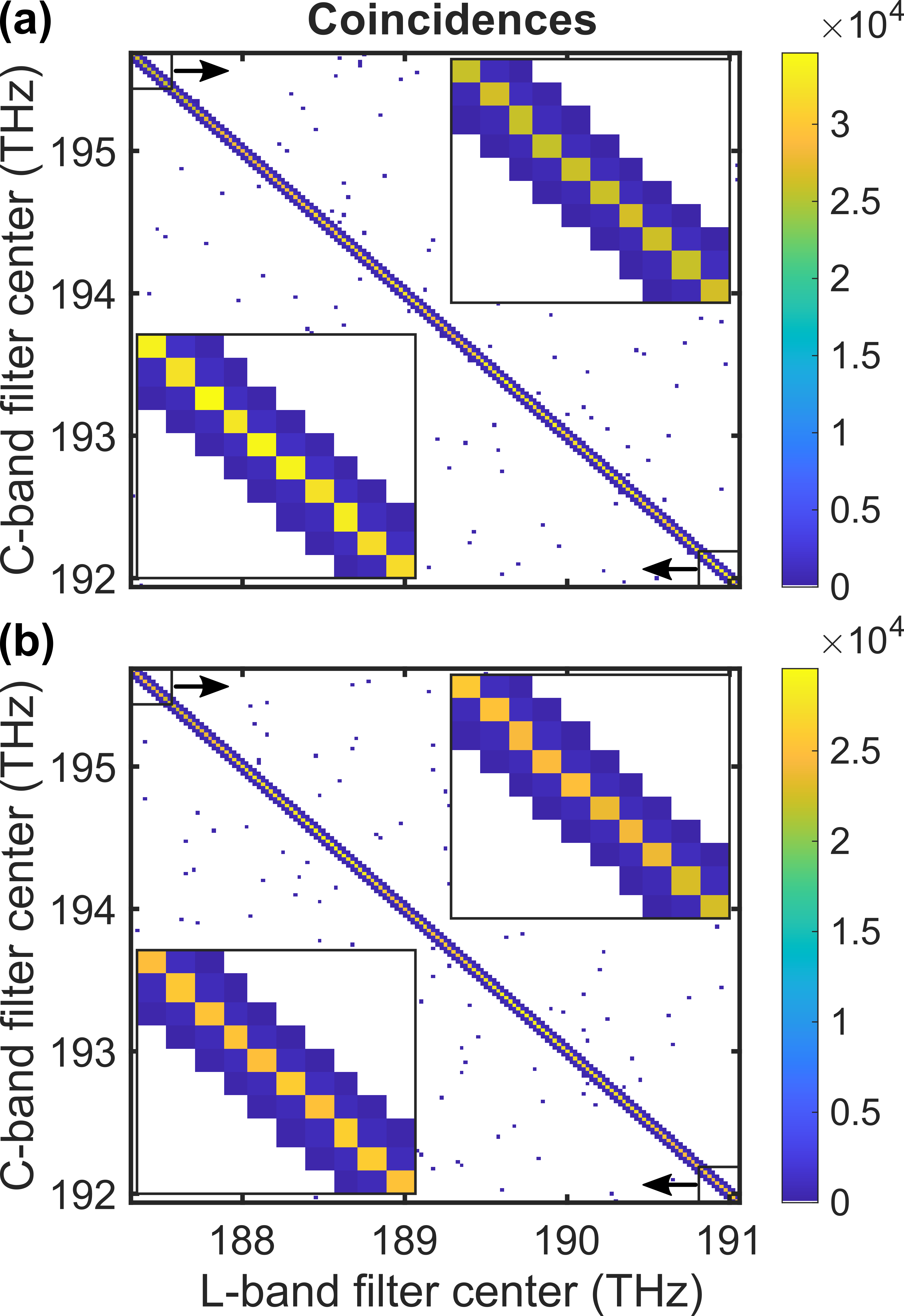}
	\caption{Partial JSI measurement in the (a) forward and (b) backward directions plotted as a function of signal (C-band) and idler (L-band) channel numbers. A 1~ns coincidence window  and 1~s integration time are used to obtain each point. The insets highlight the first and last $10\times10$ grids.}
	\label{F_JSI_col}
\end{figure}

We first characterize our entanglement source by measuring the joint spectral intensity (JSI) by removing the fiber PBS and polarization analyzers in Fig.~\ref{fig1png}, pumping the waveguide unidirectionally, and raster scanning the C- and L-band channels with the WSSs. %
The JSIs in Fig.~\ref{F_JSI_col} were obtained at a pump power of 2~mW. %
Because entanglement distribution leverages energy-matched channels only,  %
we focus our measurements on the important center region of the JSI composed of the energy-matched diagonal and two upper and two lower sidebands, sampling the remainder of the JSI with one additional random point per row. %
Both directions reveal strong correlations: the mean and standard deviation of the coincidences-to-accidentals ratios (CARs) for the forward (backward) direction are $100.0\pm 0.9$ ($113.8\pm 0.9$) on the diagonal points, $3.40\pm 0.04$ ($3.74\pm 0.04$) for the nearest sidebands and $1.044\pm 0.007$ ($1.044\pm 0.007$) for the second-order sidebands and random points. %
The small first-order sideband correlations are to be expected from the nonzero filter rolloff of the WSS passbands; other than introducing a small background into adjacent channels, this effect presents no major problems for entanglement distribution. And with CAR$\sim$1 for all second-order sidebands and beyond---indicating only accidental coincidences for these channel pairings---our results confirm accurate filter alignment and high extinction in the paired WSSs.

We characterize the quality of polarization entanglement for the 150~channel pairs by performing complete QST for each channel using the full setup of Fig.~\ref{fig1png}. %
There are three main components in each polarization analyzer: a quarter-wave plate (QWP), a half-wave plate (HWP), and a PBS, with the waveplates  controlled by a set of motion controllers. %
To optimize the pump power splitting, we first compensate for the random birefringence effects induced by the single-mode optical fiber by inserting a polarization-maintaining circulator in the Sagnac loop---permitting only one direction to pump the source---and adjust the fiber PCs before each analyzer to minimize the counts measured in $\ket{VV}$; this ensures that each direction in the loop is uniquely identified with the $H/V$ bases at each receiver. %
Removing the circulator and returning to the setup of Fig.~\ref{fig1png}, we then tune the pump's PC to balance %
the coincidences in $\ket{HH}$ and $\ket{VV}$. %
After this one-time setup, we proceed to 
QST measurements. Again, we program the C-band (L-band) WSS to send a signal (idler) to the output ports leading to the polarization analyzers. Then we collect 36~measurements for each channel for all combinations of the rectilinear $H/V$, diagonal $D/A$, and circular $R/L$ basis states. The two qubit polarization state, along with uncertainty, is inferred through Bayesian QST~\cite{Blume2010, Lukens2020b} using a Bures prior and Poissonian likelihood as outlined in \cite{Lu2021}.

\begin{figure}[tb!]
	\centering
    \includegraphics[width=\columnwidth]{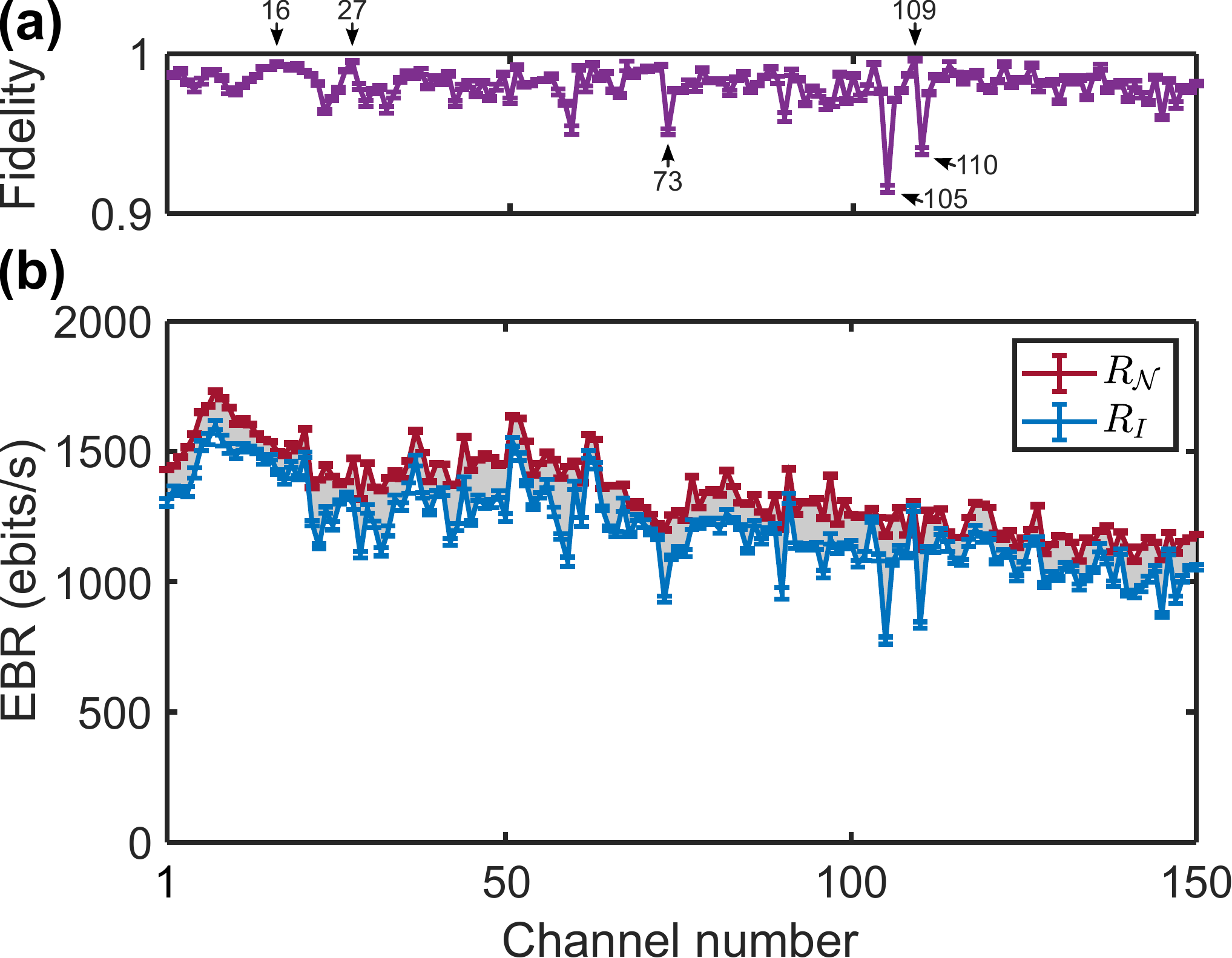}
	\caption{(a)~Bayesian state fidelities of all 150 channels. Arrows above (below) the curve indicate the three channels with the highest (lowest) fidelity. (b)~Distillable entanglement rate $R_D$ for each channel is bounded above by the log-negativity rate $R_\mathcal{N}$ %
	and below by the coherent information rate
	$R_I$. %
	}
	\label{F_EBR}
\end{figure}

Irrespective of $H/V$ alignment procedure, an additional phase shift remains in the polarization rotation from source to detector
that is a priori unknown.  Rather than finding and compensating directly, full state tomography allows us to back out these empirical rotations \emph{after} state estimation; specifically, for each channel we find and apply the rotation $U_A \otimes U_B$ which maximizes the overlap of the Bayesian mean density matrix $\rho_B$ with the ideal Bell state $\ket{\Phi^+}$. As a tensor product of local unitaries, this operation leaves all entanglement metrics invariant. Moreover, the fidelity found in this way is precisely the fully entangled fraction, a valuable quantity in its own right that sets useful bounds in quantum communications protocols such as teleportation, dense coding, and entanglement swapping~\cite{Grondalski2002}.

To quantify the rate of useful entanglement generated, we consider entangled bit rate (EBR) as defined in terms of distillable entanglement $E_D$: the  maximal asymptotic rate of Bell pair production per received state using only local operations and classical communications~\cite{Bennett1996}. Although difficult to calculate directly, $E_D$ is bounded above by log-negativity $E_\mathcal{N}$~\cite{Vidal2002, Plenio2005} and below by coherent information ($I_{A\rightarrow B}$ or $I_{B\rightarrow A}$, where the arrow denotes the direction of one-way classical communication between receiver $A$ and $B$)~\cite{Devetak2005}. Both quantities are readily computable from the density matrix; by multiplying by the observed coincidence rate $R_\text{coinc}$, we can therefore define the EBR measures $R_\mathcal{N}=R_\text{coinc}E_\mathcal{N}$ and $R_I=R_\text{coinc} \max\{I_{A\rightarrow B},I_{B\rightarrow A}\}$ with units of ebits/s, such that the distillable entanglement rate $R_D\in[R_I,R_\mathcal{N}]$. The inclusion of the lower bound $R_I$ improves upon our previous EBR analyses for entanglement distribution which focused on the upper bound $R_\mathcal{N}$ only~\cite{Alshowkan2021, Alshowkan2022b}.

The fidelity and EBR of each channel are shown in Fig.~\ref{F_EBR}(a) and (b) respectively, each obtained with a 10~s integration time per point and 1~ns coincidence window at a total pump power of 1.6~mW across the two outputs of the fiber PBS. The channels show extremely high fidelities: an average of 0.98 with standard deviation of 0.01. Slight variations in the coincidence rate lead to the slow downward trend in EBR with channel number.
The curves $R_\mathcal{N}$ and $R_I$ in the EBR plot subtend a relatively tight region for $R_D$ for each channel, with an average of $R_D=[1210,1340]$ ebits/s, making the total distillable entanglement available %
in this configuration at least as high as 181~kebits/s. %
Note that no subtraction of accidental coincidences is performed in these calculations, nor are any system efficiencies backed out of the EBR.

\begin{figure*}[thb]
	\centering
    \includegraphics[width= 5.5in]{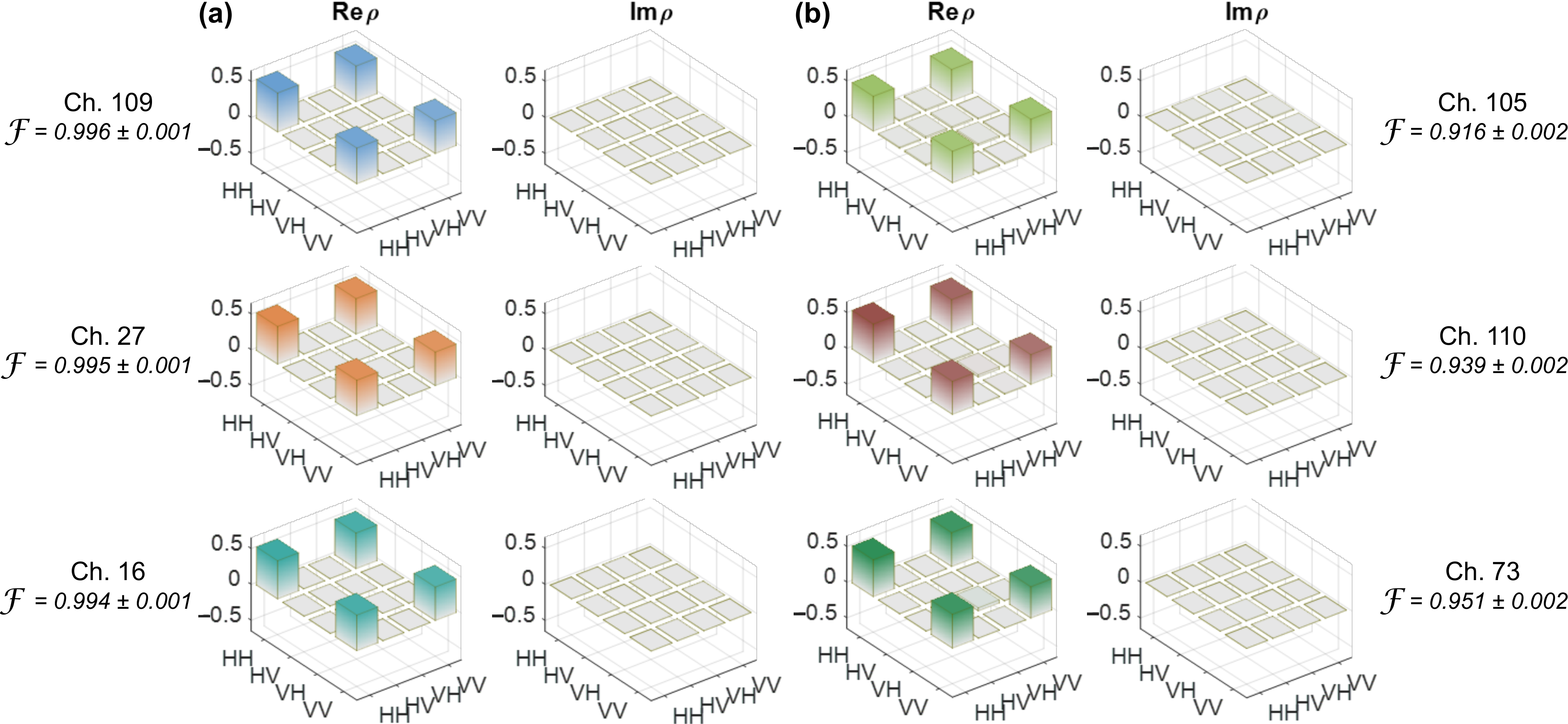}
	\caption{Density matrices of the three channels with (a)~highest and (b)~lowest fidelity as estimated by Bayesian tomography.}
	\label{density_double}
\end{figure*}

Figure~\ref{density_double} plots the (rotated) Bayesian mean density matrix $\rho_B$ for the channels with the (a) three highest and (b) three lowest fidelities. All matrices show excellent agreement with an ideal $\ket{\Phi^+}$ Bell state, confirming visually the entanglement quantified in Fig.~\ref{F_EBR}. The level of variation that is exhibited results from what we believe is the main limiting factor in our current configuration: the C-band fiber-based PBS. Beyond the support of multiple modes for the $\sim$780~nm pump laser, we have observed an excess insertion loss of 16~dB at this wavelength, as well as $\sim$20\% power fluctuations in the output ports on the scale of $\sim$15~min. Our findings contrast with the much lower $\sim$3~dB excess loss noted for the similar PBS and configuration in \cite{Vergyris2017}; because 780~nm falls so far outside of the designed PBS operating wavelength, it is unsurprising that performance could vary widely and unpredictably between specific samples. To improve pump throughput, a free-space dichroic PBS could be used; alternatively, if sufficient power in the telecom band were available, one could consider cascaded second-harmonic generation and SPDC in the same waveguide~\cite{Arahira2011}, followed by an aggressive notch filter to remove the unconverted light. %

Irrespective of these interesting research directions, our current ultrabroadband source offers high-fidelity entanglement channels capable of supporting a variety of quantum communications protocols. The WSS technology in particular enables the adaptive aggregation and assignment of multiple channels to serve users on demand with bandwidth commensurate to their needs. %
Simultaneous support of 150 pairs of users is limited in the current setup by the output port count on the specific WSSs (9 for the C-band, 20 for the L-band). Yet there have been strong advances in WSS technology in recent years~\cite{Ma2020}, enabling  narrower spacings, more output ports, and better channel isolation, so that provisioning hundreds of users should be straightforward with relatively few devices. In addition, combining the manipulation and measurement techniques developed for frequency-bin quantum information processing~\cite{Lu2018b} with previous characterization approaches~\cite{Barreiro2005}, we can demonstrate that our source is indeed hyperentangled in both polarization and frequency DoFs, offering potential in quantum communication protocols such as superdense coding~\cite{Barreiro2008}.

Since EBR is not additive, %
flex-grid performance cannot be predicted solely by combining the results in Fig.~\ref{F_EBR}, but requires an appropriate noise model such as that for optimal flex-grid allocation introduced in~\cite{Alnas2022}. Importantly, this model can also accommodate standard impairments in deployed fiber like polarization-mode dispersion and polarization-dependent loss. Consequently, detailed polarization characterization will likely prove a critical component of future quantum networks---information which not only will improve spectrum allocation decisions, but also inform analog techniques for direct, and even nonlocal, compensation~\cite{Shtaif2011,Jones2018,Kirby2019}.

\end{document}